 \documentclass[final,5p,times,twocolumn,authoryear]{elsarticle}

\usepackage{amssymb}
\usepackage{lipsum}
\usepackage{xcolor}

\usepackage{lineno}

\journal{Nuclear Physics B}

\begin{document}

\begin{frontmatter}



\title{A Study of Afterglow Signatures in NaI and CsI Scintillator Modules for the Background and Transient Observer Instrument on COSI}


\author[a,b]{Hannah Gulick}
\author[c,d]{Hiroki Yoneda}
\author[e,f]{Tadayuki Takahashi}
\author[a,b]{Claire Chen}
\author[g]{Kazuhiro Nakazawa}
\author[b,e,f]{Shunsaku Nagasawa}
\author[h]{Mii Ando}
\author[h]{Keigo Okuma}
\author[b]{Alyson Joens}
\author[b]{Samer Al Nussirat}
\author[i]{Yasuyuki Shimizu}
\author[i]{Kaito Fujisawa}
\author[j]{Takayoshi Kohmura}
\author[k]{Kouichi Hagino}
\author[l]{Hisashi Kitamura}
\author[b]{Andreas Zoglauer}
\author[b]{Juan Carlos Martinez Oliveros}
\author[b]{John A. Tomsick}

\affiliation[a]{organization={University of California, Berkeley, Department of Astronomy},
            city={Berkeley},
            postcode={94720}, 
            state={CA},
            country={USA}}
\affiliation[b]{organization={Space Sciences Laboratory, University of California},
            addressline={7 Gauss Way}, 
            city={Berkeley},
            postcode={94720-7450}, 
            state={CA},
            country={USA}}
\affiliation[c]{organization={Julius-Maximilians-Universität Würzburg, Fakultät für Physik und Astronomie, Institut für Theoretische Physik und Astrophysik, Lehrstuhl für Astronomie},
            addressline = {Emil-Fischer-Str. 31},
            city={D-97074 Würzburg},
            country={Germany}}
\affiliation[d]{organization={RIKEN Nishina Center},
            addressline={2-1 Hirosawa}, 
            city={Wako},
            state={Saitama 351-0198},
            country={Japan}}
\affiliation[e]{organization={Kavli Institute for the Physics and Mathematics of the Universe (WPI), University of Tokyo},
            city={Kashiwa}, 
            state={Chiba 277-8583},
            country={Japan}}
\affiliation[f]{organization={University of Tokyo, Department of Physics}, 
            city={Bunkyo}, 
            state={Tokyo 113-0033},
            country={Japan}}
\affiliation[g]{organization={KMI, Nagoya University}, 
            city={Nagoya}, 
            state={Aichi},
            country={Japan}}
\affiliation[h]{organization={Nagoya University, Department of Physics}, 
            city={Nagoya}, 
            state={Aichi},
            country={Japan}}
\affiliation[i]{organization={Tokyo University of Science}, 
            city={Noda}, 
            state={Chiba 278-8510},
            country={Japan}}
\affiliation[j]{organization={Department of Physics and Astronomy, Faculty of Science and Technology, Tokyo University of Science}}
\affiliation[k]{organization={The University of Tokyo, Department of Physics}, 
            city={Bunkyo}, 
            state={Tokyo 113-0033},
            country={Japan}}
\affiliation[i]{organization={National Institutes for Quantum Science and Technology}, 
            city={Inage-ku}, 
            state={Chiba 263-8555},
            country={Japan}}

\begin{abstract}
We present measurements of the afterglow signatures in NaI(Tl) and CsI(Tl) detector modules as part of the Background and Transient Observer (BTO) mission detector trade-study. BTO is a NASA Student Collaboration Project flying on the Compton Spectrometer and Imager (COSI) Small Explorer mission in 2027. The detectors utilized in this study are cylindrical in shape with a height and diameter of 5.1 cm and were read out by silicon photomultipliers (SiPMs). We conducted a radiation campaign at the HIMAC accelerator in Japan where the scintillators were irradiated with a 230 MeV/u helium beam (He beam) and 350 MeV/u carbon beam (C beam). We find that both the CsI and NaI scintillators exhibit afterglow signatures when irradiated with the C and He beams. The CsI crystal exhibits a stronger afterglow intensity with afterglow pulses occurring for an average 2.40 ms for C and 0.9 ms for He after the initial particle pulse. The duration of afterglow pulses in CsI is 8.6$\times$ and 5.6$\times$ the afterglow signal duration in NaI for C and He (0.28 ms and 0.16 ms, respectively). Although CsI has advantages such as a higher light yield and radiation hardness, the stronger afterglows in the CsI detector increase the complexity of the electronics and lead to a $\sim$7$\times$ larger dead time per afterglow event or a $\sim$3$\times$ higher energy threshold value. We use the measured dead times to predict the amount of observing time lost to afterglow-inducing events for an instrument like BTO in low Earth orbit. We simulate the background rates in a BTO-like orbit and find a total value of 114 counts/s for the full two-detector system. Based on the particle energies in the HIMAC experiment, we then determine that an event with sufficient energy to produce an afterglow signal occurs once every $\sim$70 s and $\sim$1.4 s in NaI and CsI detectors, respectively. Thus, we conclude that NaI is the better choice for the BTO mission.
\end{abstract}



\begin{keyword}
Gamma-ray scintillators \sep Scintillator Afterglow \sep All-sky gamma-ray survey \sep time-domain astrophysics


\end{keyword}

\end{frontmatter}




\section{Introduction}
\label{sec:intro}

Afterglow, in the context of scintillators, is a phenomenon during which light emission in the crystal continues after a high-amplitude signal. Afterglow emission occurs when a delayed charge carrier is thermally released and recombines at a luminescence center causing a pulse  \citep{Koppert2018, Lecoq2020, Farukhi1982}. Several factors can influence the delayed recombination of charge carriers, including structural defects or impurities in the crystal and temperature \citep{ALFASSI2009585}. The delayed pulses are a concern for high-energy astrophysics instruments for several reasons. In space-based instruments, the energetic particles in Earth's radiation environment provide a continuous source of scintillation and afterglow light emission. These particles induce phosphorescence in scintillators, thus producing an effective degradation in a detector's energy resolution \citep{Dilillo2022}. Additionally, the delayed afterglow pulses can trigger as false $\gamma$-ray events in the detector electronics. Therefore, the afterglow signature and duration for a given scintillator material must be well understood to minimize false triggers.

The work presented in this paper was motivated by a detector trade-study conducted for the Background and Transient Observer (BTO). BTO is the Student Collaboration Project that will fly on the Compton Spectrometer and Imager (COSI) Small Explorer Mission (SMEX) \citep{Tomsick2023} to study $\gamma$-ray transients and monitor the soft $\gamma$-ray background in the 30 keV to 2 MeV range. The BTO detector trade-study includes in-depth analyses of two inorganic crystals ---NaI(Tl) and CsI(Tl)--- as read out with silicon photomultipliers (SiPMs) and corresponding electronics. Both scintillator types have space heritage spanning several decades, high light output when exposed to ionizing radiation, long-term stability and durability, and a low-cost baseline due to commercial availability \citep{cryst12111517, Farukhi1982}. However, while both crystals are generally cost effective, obtaining a high-purity crystal with minimal afterglow-inducing defects and impurities rapidly drives up the price.

In this paper, we compare afterglow signatures in waveforms taken with the Scionix NaI(Tl) and CsI(Tl) 51B51/SiP-E3-X cylindrical detector modules with SiPM readouts (see Table \ref{tab:detectors} and Figure \ref{fig:detector}). The Scionix packages have a built-in, proprietary amplifier circuit, therefore we assess the afterglow signal in the combined output from the crystal, SiPM, and amplifier. Regardless of the built-in shaping parameters, afterglows are long, enduring signals intrinsic to the scintillator that cannot be neglected when designing $\gamma$-ray missions. This experiment quantifies the severity of afterglows in the Scionix NaI and CsI detectors in order to determine the time period after a heavy ion hit which is lost to afterglows. This work is applied directly to the BTO mission but is also more generally useful to space-based missions using NaI or CsI with SiPM readouts.

Typical afterglow signatures were reported to reach fractions of $\sim$3--15\% after 1--3 ms in NaI(Tl) detectors for X-ray pulses \citep{Lecoq2020, Koppert2018, Farukhi1982}. Studies on the CsI(Tl) detectors report high enough afterglow fractions in high-energy irradiation to prevent its use in most astrophysical and medical applications \citep{ALFASSI2009585, Lecoq2020, Farukhi1982}. Historically, afterglow studies in the $\gamma$-ray regime were almost exclusively conducted with photomultiplier tubes (PMTs) or photodiode readouts. However, the newer SiPM technology is becoming popular in space-based applications where their lower weight and volume (compared to a PMT) makes them advantageous. Therefore, a comprehensive study of afterglow in systems utilizing SiPMs is necessary for BTO and other future $\gamma$-ray missions.


To measure and constrain the afterglow signatures in NaI(Tl) and CsI(Tl) scintillators, we conducted an irradiation campaign at the Heavy Ion Medical Accelerator in Chiba (HIMAC). Both scintillator materials were irradiated with a 230 MeV/u (920 MeV) He beam and a 350 MeV/u (4.2 GeV) C beam. This paper describes the measurements, analysis, and afterglow results in Section \ref{sect:afterglow}. Section \ref{sect:BTO} further describes the BTO instrument with detector trade-study results in Section \ref{subsect:detectors} and simulations of the background and afterglow inducing particle rates in the BTO orbit.

\section{Measuring Afterglow Signatures in Waveform Data}
\label{sect:afterglow}

\subsection{Setup and Measurements}
\label{subsect:obs}

\begin{figure}
\begin{center}
\begin{tabular}{c}
\includegraphics[height=4.5cm]{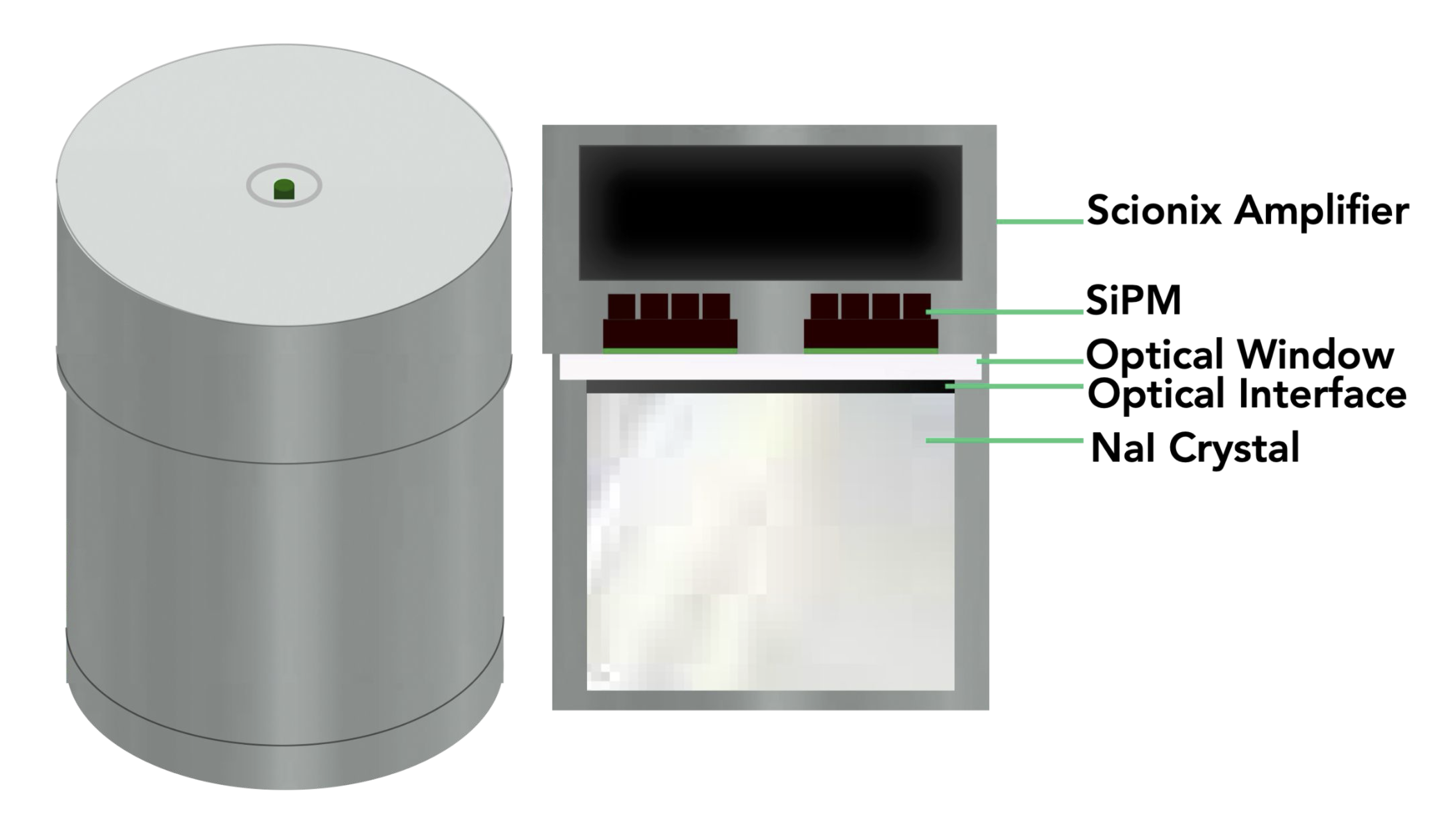}
\end{tabular}
\end{center}
\caption 
{ \label{fig:detector}
The Scionix cylindrical detector schematic. The detector contains an NaI or CsI crystal with a 5.1 cm diameter and 5.1 cm height. The aluminum case is 7.53 cm tall with a 5.92 cm diameter at the top (thickest point) and 5.7 cm at the bottom. The crystal is readout by SiPMs through an optical interface and window. The SiPM readouts are then shaped by a Scionix designed amplifier. The crystal is covered in a reflector (stretched teflon) and hermetically sealed. The detector, window, SiPMs, and electronics are housed in a 1.25 mm thick body of aluminum.} 
\end{figure}

We took measurements at the HIMAC facility in Japan using both an NaI and CsI based detector module with SiPM readouts. The crystals were packaged by Scionix and cylindrical in shape with a diameter and height of 5.1 cm. The detector modules, shown in Figure \ref{fig:detector}, included a hermetically sealed scintillator (NaI or CsI) followed by an optical window, ArrayJ-60035-4P-PCB SiPM, and a built-in amplifier specific to Scionix detectors. The SiPM is PCB mounted with a high ($\sim$90\%) fill-factor and a peak wavelength at 420 nm, dark current of $\sim$7.5 $\mu$A, and rise time of 250 ps. The scintillator and electronics were housed in a 0.125 cm thick aluminum casing. Table \ref{tab:detectors} includes more information on the Scionix detector specifications.

\begin{table}[ht]
\caption{Scionix Detector Specifications} 
\label{tab:detectors}
\begin{center}       
\begin{tabular}{l|l|l}
\rule[-1ex]{0pt}{3.5ex} Specs & CsI & NaI \\
\hline\hline
\rule[-1ex]{0pt}{3.5ex}  Pulse rise time ($\mu$s) &  1.2 & 0.56 \\
\rule[-1ex]{0pt}{3.5ex}  Pulse fall time ($\mu$s) &  2.6 & 1.4 \\
\rule[-1ex]{0pt}{3.5ex}  Energy resolution (FWHM @ 662 keV) &  $<$8\% & $<$7.5\%\\
\rule[-1ex]{0pt}{3.5ex}  Noise level (keV) &  $<$20 & $<$15\\
\rule[-1ex]{0pt}{3.5ex} Gain (mV/MeV) & 300 & $\sim$450 \\
\rule[-1ex]{0pt}{3.5ex} 30.85 keV Peak Position (ADU) & 10.6 & 14.1 \\
\end{tabular}
\end{center}
\end{table}

The scintillators were excited with particle radiation at the HIMAC accelerator on 2022 June 16–18. This study included four experimental setups:

\begin{itemize}
  \item Setup 1: CsI detector irradiated by 350 MeV/u C beam
  \item Setup 2: NaI detector irradiated by 350 MeV/u C beam
  \item Setup 3: CsI detector irradiated by 230 MeV/u He beam
  \item Setup 4: NaI detector irradiated by 230 MeV/u He beam
\end{itemize}

\begin{figure}
\begin{center}
\begin{tabular}{c}
\includegraphics[height=6.5cm]{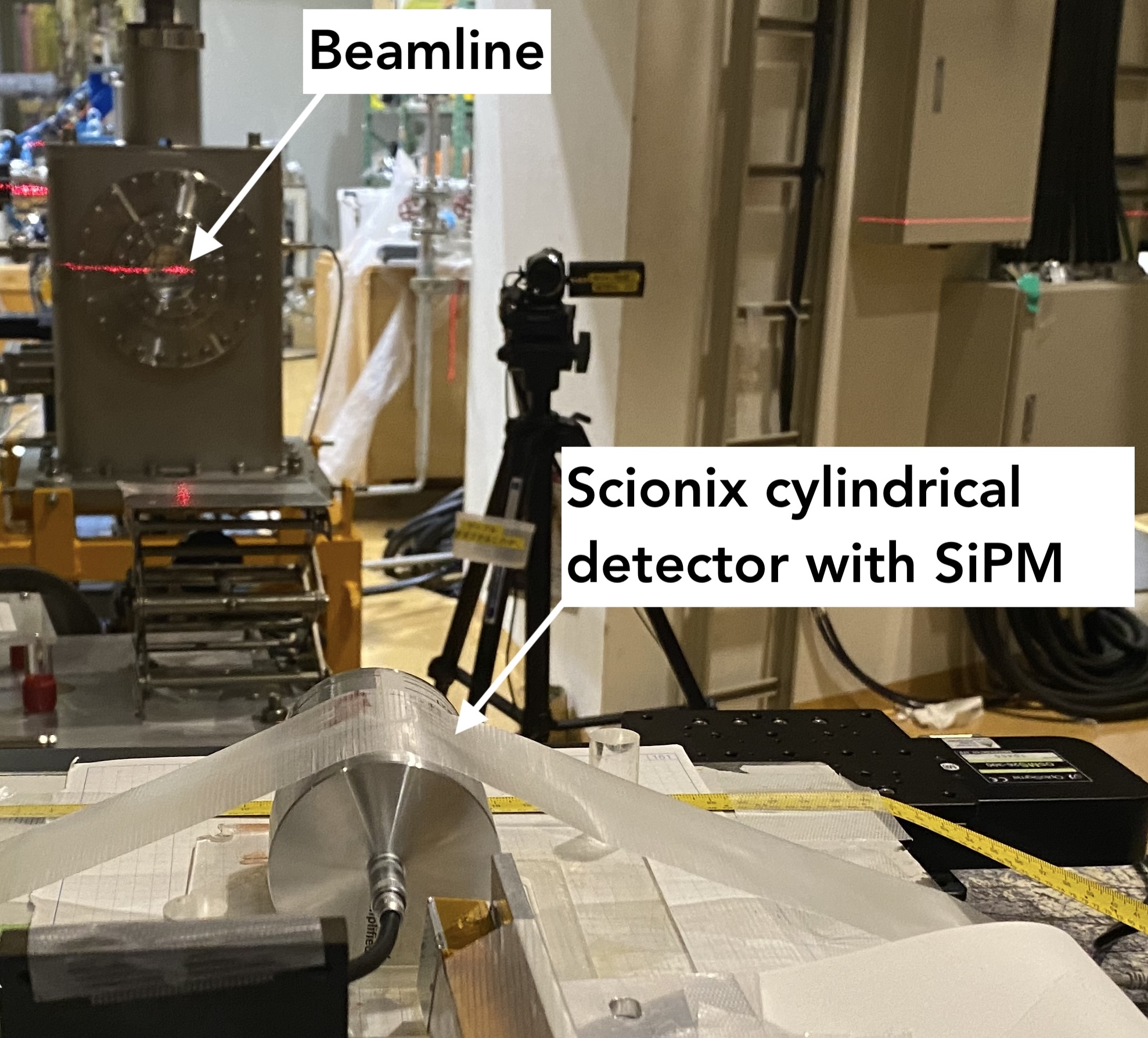}
\end{tabular}
\end{center}
\caption 
{ \label{fig:HIMAC_setup}
The HIMAC experimental setup. The Scionix detector (NaI or CsI) is labeled and appears as the silver, cylindrical tube taped down to the table. The beamline is labeled in the back of the photo and has a red alignment laser bisecting the center of the beam. 
} 
\end{figure}

\noindent Each detector was operated with an input voltage of $\pm$8 V and irradiated for $\sim$20 minutes in both beamlines at the given detector orientation (with respect to the beam). Figure \ref{fig:HIMAC_setup} shows the experimental setup, with the accelerator beam and Scionix detector labeled. 


As shown in Figure \ref{fig:HIMAC_setup}, the cylindrical detectors were aligned so the beam was along the detector's long axis. The particles entered the circular face of the detector on the non-electronics side and passed directly into the NaI or CsI detectors. Aligning the beam along the crystal's longest axis allowed for the most energy to be deposited within the crystal resulting in the worst-case-scenario for afterglow signatures in the Scionix detectors. However, it is important to note that even with the beam aligned along the crystal's longest axis, the kinetic energy of both the He and C particles results in the a particle range longer than the detectors length---i.e. the particles are not fully absorbed and only a fraction of the energy is deposited in the detector. For the CsI detector, the average energy deposited by the He and C beam is 325 MeV and 2 GeV, respectively. In the NaI detector the average energy deposited by the He and C particles is 275 MeV and 1.65 GeV, respectively. Additionally, by pointing the crystal end of the detector toward the beam, a given particle's energy was not depleted through scatters in the electronics, and the electronics were shielded and protected from the highest doses of radiation.

\begin{figure*}
\begin{center}
\begin{tabular}{c}
\includegraphics[width=16cm]{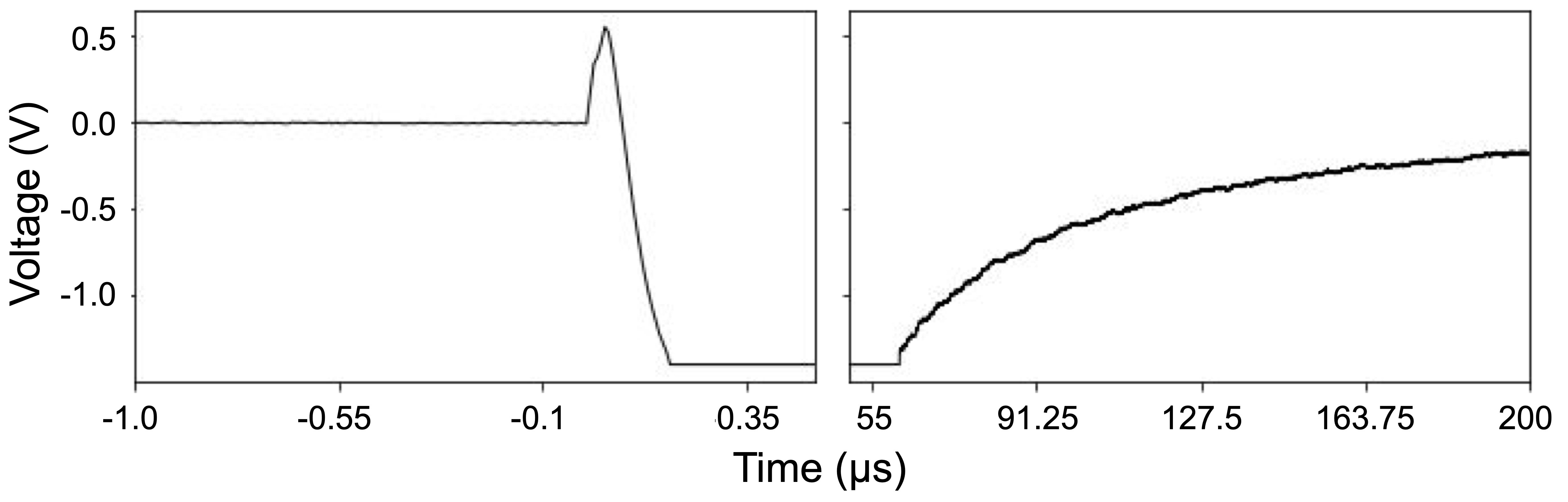}
\end{tabular}
\end{center}
\caption 
{ \label{fig:waveform_ex}
An example waveform taken with the CsI detector and the C beam. The time-axis values are chosen to highlight the pulse's features including the [left] positive spike in the waveform voltage at the start of the saturated pulse at zero seconds and [right] the pulse's return to baseline with afterglow signatures starting at $\sim$0.057 ms. The afterglow features are small pulses in the waveform data, and at this scale, appear as a broadening in the voltage distribution (i.e. thicker line) at times after $\sim$0.057 ms (right plot) when compared with the pre-pulse, baseline voltage distribution at times below -0.001 ms (left plot). The time-axis is not shown from 0.0005 ms to 0.05 ms since the detector is saturated and the voltage value stays constant at -1.4 V, as shown in this figure at the highest and lowest timescales in the left and right subplots, respectively. } 
\end{figure*}

Waveform data was taken during each experimental setup using a PicoScope 5444D with a 200 MHz dB bandwidth, 4 potential readout channels, and a sample rate of 125 MS/s. Using the PicoScope6 Alarm functionality, the PicoSope was set to auto-trigger and auto-save a waveform for each down-going DC signal that crossed a --0.1 V threshold. The expected maximum afterglow duration time was several milliseconds so the scope's time-axis was set to have 2 ms/div to ensure at least one pulse and complete afterglow were included in each waveform. The voltage-axis was set to $\pm$1.5 V to capture the full saturated or unsaturated pulse shape. The samples were taken with 16 bits and a resolution of 1.25 mega-samples per waveform for the C beam data. During the He beam test, a second PicoScope channel was used to monitor a shaped output so the data were taken with 15 bits and a 2.5 mega-sample resolution. The resolution for both the carbon and helium data is fully sufficient to capture the afterglow signatures in the data. Approximately 100 waveforms were taken for each detector in each experimental setup. Figure \ref{fig:waveform_ex} shows an example of the waveform data taken with the CsI detector in the C beam.

\subsection{Fitting the Waveform Data}
\label{subsect:analysis}

\begin{figure*}
  \centering
  \begin{tabular}{@{}c@{}}
    \includegraphics[width=.8\linewidth,height=105pt]{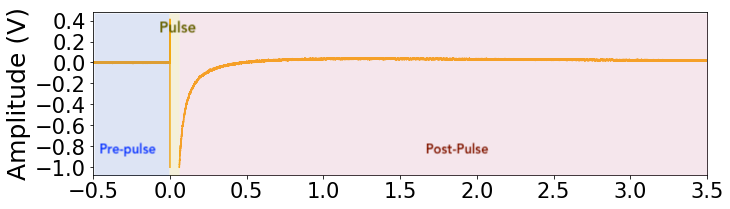} \\[\abovecaptionskip]
  \end{tabular}

  \begin{tabular}{@{}c@{}}
    \includegraphics[width=.8\linewidth,height=105pt]{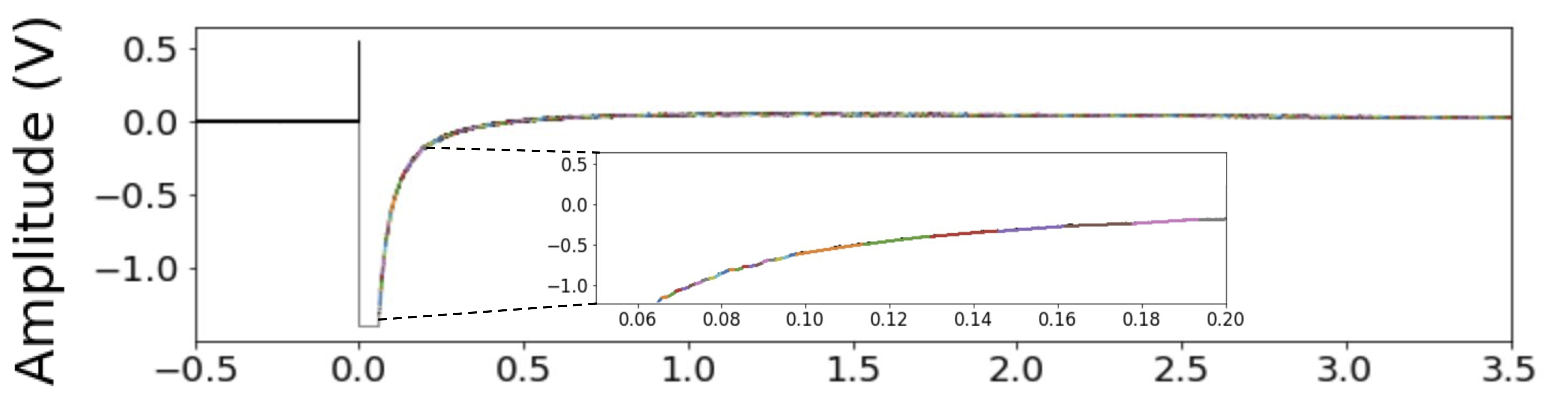} \\[\abovecaptionskip]
  \end{tabular}

  \begin{tabular}{@{}c@{}}
    \includegraphics[width=.8\linewidth,height=130pt]{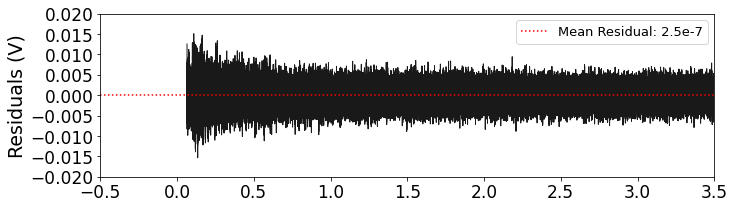} \\[\abovecaptionskip]
  \end{tabular}

  \begin{tabular}{@{}c@{}}
    \includegraphics[width=.8\linewidth,height=150pt]{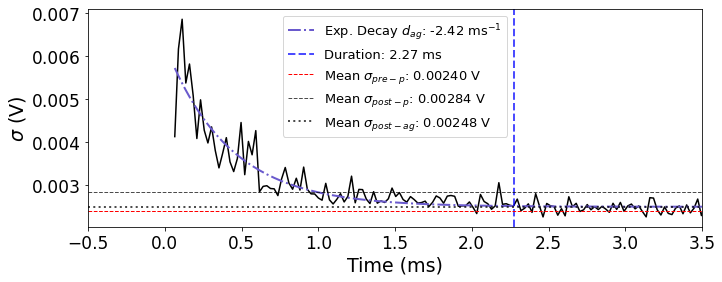} \\[\abovecaptionskip]
  \end{tabular}

  \caption{(Top) An example of a waveform taken with the CsI detector with SiPM readout when irradiated with the C beam. The region classification used in this study and discussed more in Section \ref{subsect:analysis} is overplotted as shaded regions with blue being the pre-pulse, yellow the pulse, and red the post-pulse regions. (Middle-top) The series of exponential fits overplotted on the post-pulse waveform. Each colored line represents a different exponential fit as described in Section \ref{subsect:analysis}. The subplot contains a zoomed-in view of the immediate post-pulse region and has the same axis units as the full figure. (Middle-bottom) The residuals as a function of time after the waveform in the top panel was fit to remove general trends from the data. By removing any trends, the afterglow signature can be mostly isolated (see Section \ref{subsect:analysis}) and appears here as the wider spread in voltage starting at 0.0 ms. (Bottom) A distribution of the voltage spread in the residuals from the middle panel. The voltage spread is measured over bins of 24 $\mu$s and plotted against time. The mean pre-pulse spread, plotted as the red dashed horizontal line, is calculated as the average baseline voltage spread for 16 $\mu$s time bins in the pre-pulse data (defined in Section \ref{subsect:analysis}) and is taken to be the baseline voltage spread without afterglow. The mean post-pulse spread, plotted as the black dashed line, shows the average spread in the post-pulse voltage residuals and is several mV larger than the pre-pulse spread further emphasizing the presence of afterglow. The measured afterglow duration is shown as a vertical dashed line, and the mean post-afterglow spread, plotted as a black dotted line, shows the average voltage spread of the post-pulse data at timescales after this duration. The error bars, as explained in Section \ref{subsect:analysis}, are included in the bottom panel and are on the order of $10^{-7}$ V. Thus, they are not visible in the figure.} \label{fig:analysis}
\end{figure*}

The waveform data around a pulse is separated into three regimes in this paper: the pre-pulse, pulse, and post-pulse regions (see Figure \ref{fig:analysis}). The ``pre-pulse region" includes all data points from --0.5 ms to the beginning of the pulse at zero seconds. The beginning of the pulse is different in saturated versus unsaturated waveform data. As shown in the left subplot of Figure \ref{fig:waveform_ex}, in saturated instances the pulse appears as a rapid, positive spike in the waveform voltage which then drops rapidly down to the saturation limit of --1.4 V in 0.18 $\mu$s.
Both the positive spike and saturation limit are characteristics of the Scionix electronics package---the saturation limit being determined by the detector's preset SiPM gain and the positive spike being a feature of the proprietary built-in amplifier circuit. After dropping in voltage, the waveform keeps a constant value of --1.4 V for the saturation duration and then follows a roughly exponential increase back to the pre-pulse baseline waveform voltage. The saturation duration increases with the energy of the particle, with 20.4 $\mu$s for C beam and 14.6 $\mu$s for He beam in the NaI detector, and 61.2 $\mu$s for C beam and 26.6 $\mu$s for He beam for the CsI detector. In the unsaturated waveform (minimum voltage $>$ --1.4 V), the pulse will appear as a sharp drop in voltage followed immediately by the roughly exponential increase back to the baseline voltage. Therefore, the ``pulse region" includes all data from the beginning of the event---marked by either a voltage spike or voltage drop---until the waveform begins its return to baseline voltages. The waveform---once increasing in voltage after the saturation period---and the following baseline is considered the ``post-pulse data."

The exponential increase in voltage which denotes the beginning of the post-pulse regime is intrinsic to the specific scintillator material and detector electronics and is different from the afterglow signature. 
An individual afterglow pulse is much smaller in both amplitude ($\sim$10 mV) and duration ($\sim \mu$seconds) than the event pulse which has a saturated amplitude of --1.4 V and a duration on the scale of a couple ms. The exact duration depends on the incoming particle's energy. Therefore, the general shape of the post-pulse waveform's return to baseline must be removed to disentangle the detector's scintillator/electronics response from the afterglow signatures. This can be done by fitting the general trend of the data in the post-pulse regime on timescales $>>$ a single afterglow event duration. Subtracting the fit from the waveform leaves a ``flat" residual (see middle panel of Figure \ref{fig:analysis}) with voltages centered around 0.0 V. The spread in this residual waveform is characteristic of the number/intensity of afterglow events at a given time step after the end of the pulse.  Also to be considered in the general post-pulse fit, are random and sinusoidal noise components. These noise components appear throughout the waveform, can be identified by eye in the pre-pulse baseline voltages, and are caused by the system's electronics and signal echoing in the 30 m cable used to connect the detector to the Picoscope and computer, respectively. These variations have an approximate timescale of 100 $\mu$s and are not consistently present throughout the waveform.

As shown in Table \ref{tab:detectors}, the NaI and CsI detectors have different gains requiring a gain correction be applied before making comparisons between waveform data. We determine the gain difference around 30 keV by observing the 30.85 keV line from a radioactive barium-133 (Ba-133) source. The same Ba-133 source was placed separately near the CsI and NaI detectors such that it produced unobscured radiation. The Ba-133 peak is then fit with a Gaussian to find the centroid value in a given detector. For the NaI and CsI detectors, the centroid is measured at 14.1 analog-to-digital units (ADU) and 10.6 ADU, respectively, revealing the gain of the NaI detector is 1.33 times higher than that of the CsI. This agrees reasonably well with the manufacturer reported gain ratio in Table \ref{tab:detectors}. To correct for the gain offset, the CsI waveform data were multiplied by a factor of 1.33.

After applying the gain correction, we isolate the afterglow signal in a given pulse by fitting and removing the general trend in the post-pulse data---i.e. the large increase in waveform voltage as it returns to baseline values after a pulse. This general trend can be seen in the top panel of Figure \ref{fig:analysis} from $\sim$0.05 ms to $\sim$1 ms where the post-pulse data increases from -1.4 V to $\sim$0.0 V. This increase in voltage is intrinsic to the detector + amplifier electronics and is not influenced by the presence of afterglow events. The middle panel of Figure \ref{fig:analysis} shows the post-pulse residual after the general trend is removed. Since the afterglow events in the data are short ($\sim \mu$second), low-amplitude ($\sim$10 mV) pulses, they appear as a temporary broadening in the waveform voltages following a high energy event. Thus, we determine the duration and amplitude of the afterglow by measuring the post-pulse waveform spread ($\sigma$) as a function of time.
Due to the complexity of energy transfer between multiple radiating centers and quenching mechanisms in scintillating materials---as well as the influences from the electronics components such as the amplifier, high-voltage supply, and high-voltage supply filter---the post-pulse data cannot be fit with a single exponential model.
Instead, we follow the practice of describing the pulse curve by a series of exponential functions with different time constants \citep{Lecoq2020}. Note that this model does not have physical meaning and is simply used to remove the intrinsic scintillation and electronics signals. An automated code was created to fit the post-pulse data with a series of exponentials allowing for modeling of both the intrinsic exponential rise to baseline, the cable noise in the post-pulse waveform, and electronics noise on timescales greater than $\sim$16 $\mu$s as determined by our time bin selection (defined below). For the first 30 $\mu$s of the post-pulse region, the voltage increase is much steeper than the subsequent increase. Therefore, the waveform data in the first 30 $\mu$s was binned into 1.5 $\mu$s bins and each bin was fit with an exponential function to fully capture the trend in the waveform. To save on computational power, the time bins were increased to 16 $\mu$s intervals after the initial 30 $\mu$s of the pulse and kept consistent out to 3.5 ms. The 16 $\mu$s time interval is approximately a sixth of the sinusoidal noise component's timescale of 100 $\mu$s, and thus, a single sinusoidal noise fluctuation is fit at six points ensuring the noise was properly modeled and removed. Using these defined time bin intervals, this resulted in 217 total automated fits---20 with 1.5 $\mu$s and 197 with 16 $\mu$s intervals. The second panel in Figure \ref{fig:analysis} shows the exponential fits overplotted on the post-pulse data from the top panel, with each color representing an single fit. The two time-bins (1.5 $\mu$s and 16 $\mu$s) are highlighted in the zoomed-in sub-figure.

Once the data were fit, the exponential models were subtracted from the waveform to create a ``flat" post-post residual centered at 0.0 V. The third panel of Figure \ref{fig:analysis} shows the residuals from the waveform in the first/second panel plotted as a function of time. The excess spread in the post-pulse residuals at low timescales is a direct consequence of afterglow signatures in the waveform. Therefore, the spread of the post-pulse residuals was measured as a function of time to determine the timescale over which afterglow events were detectable in the waveform. For a single pulse, the data were separated into subsequent bins of 24 $\mu$s. A histogram of the residual voltage values in a given 24 $\mu$s bin was then fitted with a Gaussian distribution to get the $\sigma$ of the waveform voltages at that time bin. The $\sigma$ for each bin was plotted against time to show the spread in the post-pulse waveform over time. The $\sigma$ versus time distribution was then fit with a decaying exponential with exponent $d_{ag}$ to characterize the afterglow signature's decay to baseline voltage values. The bottom panel of Figure \ref{fig:analysis} shows the time versus $\sigma$ for the single pulse residuals in the middle panel along with the exponential decay fit.

Uncertainties for the $\sigma$ values are given as the sum of the errors associated with the fits throughout the analysis process. The first uncertainty component comes from the series of 217 exponential fits. For each exponential fit segment, the mean residual value was calculated. If the fit were perfect, the residual mean should be nearly zero. As a consequence of the model being fit to the center of the waveform voltage distribution, the negative going afterglow pulses are roughly symmetric about the zero voltage axis in the residual as shown in Figure 4, third panel. Therefore, the spread as a function of time, though both positive and negative going in the residual plot, is induced by the negative going afterglow pulses. Any significant deviation from zero in the full residual voltage distribution mean value implies noise or general trending was not removed from the data and is taken to be the error associated with that data segment. These errors are then averaged over each time bin to give the uncertainty associated with the exponential fit per time bin. The second uncertainty component comes from the Gaussian fit to the voltage spread in each time bin. The fit errors are calculated from the covariance matrix for each bin.

This process---i.e. fitting the post-pulse general trend, creating fit residuals, and fitting the spread in the residuals---was repeated for a minimum of 50 pulses per testing setup. The 50 measurements were used to calculate the median afterglow's duration, amplitude, and exponential decay constant for each testing setup as discussed further in Section \ref{subsect:waveforms}.

\subsection{Afterglow Durations and Amplitudes}
\label{subsect:waveforms}

\begin{figure*}
\begin{center}
\begin{tabular}{c}
\includegraphics[height=11cm]{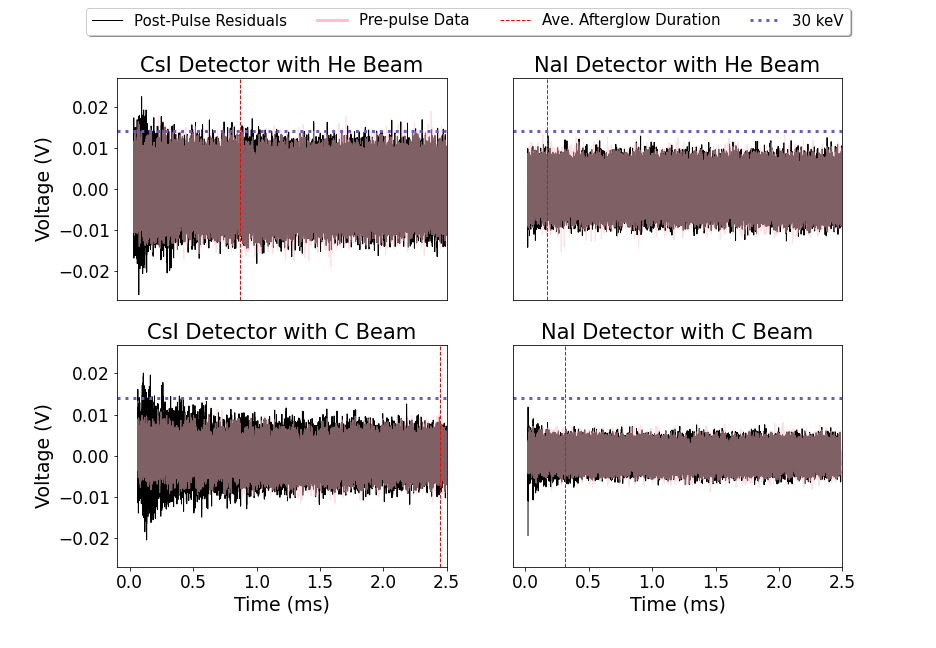}
\end{tabular}
\end{center}
\caption 
{ \label{fig:HIMAC}
The post-pulse residuals (black) for a single pulse in all four experimental setups. The residuals are found by subtracting a fit to the general trend of the post-pulse waveform from the waveform data (see Section \ref{subsect:analysis}). The afterglow signature appears as the larger spread in voltages near 0.0 seconds in each subplot. To highlight the afterglow amplitude and duration in the residual data, the pre-pulse data is overplotted in pink. The respective afterglow duration is marked with a red dashed line in each subplot. Note that the difference in the baseline voltage spread between the CsI and NaI data for a given beam (i.e. CsI with He beam and NaI with He beam) is due to the gain correction applied in Section \ref{subsect:analysis}. The difference in baseline voltage between the C and He beam for a given detector (i.e. He beam in NaI and C beam in NaI) is likely due to a difference in the ambient noise of the HIMAC facility on the different days the data were taken. The blue dashed line indicates the voltage corresponding to  30 keV for each detector.}
\end{figure*}

Figure \ref{fig:HIMAC} shows the post-pulse residuals in black for the CsI detector with He beam (top left), NaI detector with He beam (top right), CsI detector with C beam (bottom left), and NaI detector with C beam (bottom right) for a single pulse. Overplotted in a semi-transparent pink/mauve is 2.5 ms of the pre-pulse baseline data. The purpose of the pink overlay is to provide a comparison to the baseline voltage values to further highlight the spread in the post-pulse residuals caused by the presence of afterglow. The afterglow signal can be seen as the larger spread in the post-pulse residual voltages that is evident on the left end of each residual plot. This spread decreases over time, and eventually returns to the nominal baseline voltage spread, as the remaining free-electrons in the scintillator causing the afterglow spikes are captured by atoms. The afterglow signal is easily identified by eye in all cases except for the NaI detector with He beam.

The afterglow duration is taken to be the time it takes for the residual voltage $\sigma$ to reach baseline voltage values. We measure the afterglow duration using the bottom panel of Figure \ref{fig:analysis}, where the post-pulse residual voltage $\sigma$ is shown as the black solid line. The red dashed line shows the mean pre-pulse (i.e. baseline) voltage values. The black dashed line shows the mean post-pulse voltage value which is $\sim$ 0.4 mV larger than the pre-pulse mean and is indicative of afterglow events in the data. To determine the point at which the afterglow events are no longer detectable in the data, we calculate the mean voltage value over 15 consecutive time bins---or 0.36 consecutive milliseconds---in the post-pulse residual $\sigma$. When the mean over the 15 time bins is within one standard deviation of the mean pre-pulse spread ($\sigma_{pre-p}$), we define the afterglow events to be undetectable in the data. To ensure we do not average out infrequent afterglow signals toward the tail end of the post-pulse region, we add an additional constraint requiring all data points within the 0.36 ms range be within 3$\sigma$ of the mean $\sigma_{pre-p}$. Additionally, to ensure the afterglow duration is not underestimated, we require the 0.07 ms preceding the 0.36 ms range must not have an afterglow spike $>$3$\sigma$ over the mean $\sigma_{pre-p}$. We find that without this constraint, the end of the afterglow events can be defined on the declining edge of the afterglow tail or an afterglow spike, and thus the addition of this constraint more accurately encompasses the afterglow duration by providing a buffer between the last afterglow spikes and our defined return to baseline. Finally, the peak afterglow amplitude was taken to be the highest value of $\sigma$ in the $\sigma$ versus time plot (e.g. bottom panel of Figure \ref{fig:analysis}).


We find that afterglow signals are detectable in the CsI and NaI detectors when irradiated with both C and He beamlines. Table \ref{tab:results} shows the median afterglow duration, amplitude, and exponential decay profile calculated over more than 50 pulses for each of the four test cases. The parameter errors are calculated as the standard deviation over a histogram of the values measured over the $>$50 pulses. The parameter distributions for each case are symmetric. Outliers $\>$3$\sigma$ were removed from the data and usually caused by event pile-ups resulting in increased measurements of the afterglow duration.

As expected, the higher energy of the C beam results in afterglow pulses occurring over a longer timescale after each pulse when compared to the lower energy He beam. This is true for both scintillator materials and is due to the higher energy particle releasing more electrons in the scintillator material, thus increasing the chances an electron gets trapped in a crystal defect and produces an afterglow event.
Additionally, the afterglow signal is present in the CsI detector for 8.6$\times$ and 5.6$\times$ the duration of the NaI signal for the C and He beam, respectively. For the C beam, we measure an afterglow duration of 2.4 $\pm$ 0.5 ms in the CsI detector and 0.28 $\pm$ 0.07 ms in the NaI detector. For the He beam, we measure an afterglow duration of 0.9 $\pm$ 0.2 ms in the CsI detector and 0.16 $\pm$ 0.04 ms in the NaI detector. The final values and errors are calculated over 50 or more analysed pulses. Furthermore, the strongest afterglow signal in the NaI detector is weaker than the lowest energy He signal in the CsI detector, further emphasizing the CsI detector's tendency to exhibit afterglow signals. 

Furthermore, the exponential decay parameter, $d_{ag}$, is larger for the NaI detector test cases emphasizing that afterglows in this detector material decay more rapidly than in CsI. Note that the larger errors associated with $d_{ag}$ in the NaI test cases arises from the smaller afterglow signal in this detector material. With the smaller signal, there were less data points exhibiting afterglow in the spread versus binned time distribution to fit over, and the lower statistics resulted in a larger $d_{ag}$ distribution spread.

\begin{table*}[ht]
\caption{Measured Afterglow Durations and Amplitudes} 
\label{tab:results}
\begin{center}       
\begin{tabular}{l|l|l|l|l} 
\rule[-1ex]{0pt}{3.5ex} Detector & Beam & Afterglow Duration & Afterglow Amplitude & $d_{ag}$ \\
& & (ms) & (mV) & (ms$^{-1}$)\\
\hline\hline
\rule[-1ex]{0pt}{3.5ex}  CsI & C & 2.4$\pm$0.5 &  6.7$\pm$0.8 & --2.5$\pm$0.2\\
\rule[-1ex]{0pt}{3.5ex}   & He & 0.9$\pm$0.2 &  6.2$\pm$0.5 & --5.1$\pm$1\\
\rule[-1ex]{0pt}{3.5ex}  NaI & C & 0.28$\pm$0.07 &  3.3$\pm$0.3 & --12 $\pm$2\\
\rule[-1ex]{0pt}{3.5ex}   & He & 0.16$\pm$0.04 &  3.18$\pm$0.06 & --15$\pm$3\\
\end{tabular}
\end{center}
\end{table*}

\section{The Background and Transient Observer}
\label{sect:BTO}

BTO is a detector system that will fly as a student collaboration project on the larger COSI SMEX mission slated to launch in 2027. \citep{Tomsick2023}. BTO will utilize two scintillators with SiPM readouts to detect transient $\gamma$-ray sources and monitor the soft $\gamma$-ray background. 
BTO will utilize two operational modes---binned and event-by-event. Nominally, BTO will operate in binned mode, saving background information in histograms with a relative timing resolution of 0.1 ms. However, if the instrument is triggered by a $\gamma$-ray transient, BTO will also operate in event-by-event mode, saving the energy and timing information for each individual photon. The event-by-event mode retains timing information necessary to detect short events such as terrestrial $\gamma$-ray flashes and to model features in transient spectra. Binned mode data will continue to be saved even when the instrument is triggered.


The BTO system will observe in the 30 keV to 2 MeV range, thus providing spectral data at energies lower than the COSI instrument (nominal range of 0.2 MeV to 5 MeV). By extending the mission bandpass to lower energies, BTO will enable the study of potential photospheric emission components in GRB \citep{Peer2006, Ryde2010, Guiriec2011}, timing and energy characteristics of magnetar flares \citep{Peer2006, Kaspi2017}, and the origins of terrestrial $\gamma$-ray flashes \citep{Fishman1994, Smith2005, Grefenstette2009, Briggs2010, Dwyer2012}. Each BTO detector FOV will overlap with the main COSI Compton telescope FOV to provide simultaneous observations of transient events from 30 keV to 5 MeV.

Below, we discuss the implications of afterglow on the BTO system. In Section \ref{subsect:countrates}, we estimate the number of particles with energies high enough to induce afterglows in the CsI and NaI detectors. We translate this rate to the total observing time lost to afterglow events. Finally, in Section \ref{subsect:detectors}, we discuss further considerations that went into the BTO detector trade study final selection.


\subsection{Expected Afterglow Count Rates in BTO}
\label{subsect:countrates}

As part of the BTO detector trade study, we simulate the soft $\gamma$-ray background observed by the BTO detectors to estimate the rate of afterglow inducing events in CsI and NaI scintillators. In LEO, afterglow emission will be induced by particles trapped in the Van Allen radiation belts interacting with the BTO scintillators. While most of the LEO radiation background is roughly constant, areas like the South Atlantic Anomaly (SAA) and polar regions contain much higher energetic particle concentrations. Afterglow emission in these regions will be strong and result in periodic degradation in the BTO resolution and performance. Fortunately, BTO will occupy a low-inclination orbit\footnote{The current requirement for the COSI spacecraft is an orbit with $<$ 2$^\circ$ inclination.}, and therefore will have short crossing times in the SAA. We will keep BTO powered on in the SAA to measure background levels.

We utilize the Medium-Energy Gamma-ray Astronomy Library (\textbf{MEGAlib}) \citep{Zoglauer2006} to model the expected soft $\gamma$-ray background in the BTO orbit outside of the SAA. The \textbf{MEGAlib} package \textit{BackgroundGenerator} uses background models from The Space Environment Information System (SPENVIS), \cite{Ajello2008}, and \cite{Mizuno2004} to produce estimates of the background rates expected in a given orbit. Models for the trapped proton and electron fluxes, cosmic alpha fluences, and cosmic proton fluences are pulled from SPENVIS for an altitude of 550 km and inclination of 0$^\circ$ to reflect BTO's potential orbit. All values are calculated from the AP9 model at the 90th percentile. The model for the albedo photon spectrum was pulled from \cite{Ajello2008} and \cite{Mizuno2004}.

\begin{figure}
\begin{center}
\begin{tabular}{c}
\includegraphics[height=7.cm]{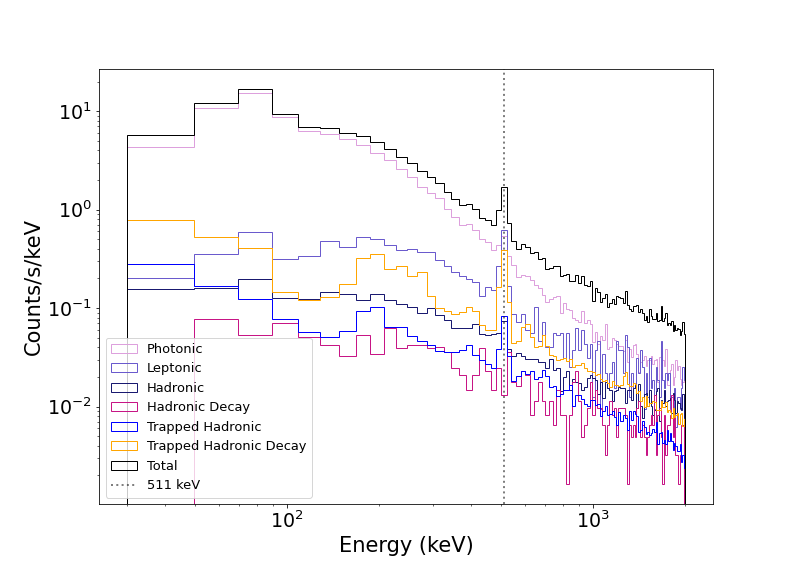}
\end{tabular}
\end{center}
\caption 
{ \label{fig:background}
The estimated background rates for the BTO system including two rectangular NaI detectors (see Section \ref{subsect:obs}). The total background rate and 511 keV peak are denoted with a black solid and dotted line, respectively.  } 
\end{figure}

The BTO instrument includes two custom NaI detectors from Scionix.
The flight-model BTO detectors will have the same scintillator volume and characteristics (see Section $\ref{subsect:obs}$) as the HIMAC cylindrical detectors, however they will be rectangular in shape to improve rough localization capabilities and mechanical mounting. The flight-model detectors contain a 3.8 cm $\times$ 3.8 cm $\times$ 7.6 cm NaI rectangular crystal hermetically sealed in a 4.5 cm $\times$ 4.5 cm $\times$ 12 cm aluminum housing \citep{Gulick2024}.
We simulate the background in the rectangular, flight-model detectors and include the full COSI mass model with representative instrument/electronics locations to ensure background scattering from the COSI materials is included in our background and afterglow calculations. The simulated background components include: photonic, leptonic, hadronic, trapped hadronic, hadronic decay, and trapped hadronic decay. Figure \ref{fig:background} shows the background rates from 30 keV to 2 MeV for the full, four-detector BTO system. The integrated background rate is calculated as the sum of the spectral counts per bin (counts/s/keV) times the bin width (in keV). We find the total integrated background rate to be 114 counts/s for two BTO detectors in low-Earth-orbit, or 57 counts/s per BTO detector.

To determine the afterglow inducing particle count rate, we consider only the integrated count rate for the hadronic and trapped hadronic components, as well as their corresponding decay components. The \textbf{MEGAlib} cosmic hadronic model includes signatures from cosmic protons, atmospheric neutrons, and alpha particles. The trapped hadronic model includes signatures from the same particles which are localized to the SAA. We then set a threshold in energy for the NaI and CsI detectors, above which all particles induce an afterglow signal. For the NaI crystal, we used a 920 MeV initial excitation energy threshold---or the total energy of the He particle in the HIMAC experiment. This value was chosen since the He afterglow signal in the NaI detector can be approximated as zero for these calculations (see Figure \ref{fig:HIMAC} and Table \ref{tab:results}). The CsI detector exhibited strong afterglow signals from both the He and C beams, which required us to estimate the afterglow inducing energy threshold. We did this by plotting the excitation energies for the He and C beam versus their corresponding afterglow durations and extrapolating to an energy which had an approximately zero millisecond afterglow duration. To allow for direct comparison with the NaI detector particle count rate, we defined the threshold energy to be the point at which the afterglow duration for the CsI detector was equal to the afterglow duration from the He beam in the NaI detector. We placed an additional constraint on the excitation energy versus afterglow duration distribution requiring that no afterglow be present at 662 keV as determined in calibration testing during the HIMAC experiment when measuring waveform data for a cesium-137 source with a peak at 662 keV. With the additional 662 keV energy constraint, we find the distribution is best fit with a second order polynomial and gives an afterglow inducing threshold of 164 MeV for the CsI detector. To check the validity of this method, we also calculate the total deposited energy necessary to produce an afterglow event in the CsI detector and compare it to results in \cite{Rau2005}. \textbf{MEGAlib} was used to calculate the total deposited energy in a 5.1 cm CsI crystal (i.e. cylindrical detector) from a single particle beam. The beam had an energy equal to the total excitation energy of the He (920 MeV) or the C (4.2 GeV) particles. We find a total deposited energy of 74.5 MeV will produce afterglows in the cylindrical CsI crystal. This value agrees with the findings of \cite{Rau2005} and falls within their expected range of  8--80 MeV. Note that while these values are calculated from the cylindrical detector tests at HIMAC, they also apply to the rectangular detectors since the afterglow inducing energy is intrinsic to the scintillator material.


We then take the integrated hadronic count rate above both energy thresholds and combine it with an assigned dead time to find the observation time lost to afterglows in both detectors.
The count rate is 0.014 particles/s (event every 74 seconds) and 0.7 particles/s (event every 1.4 seconds) for NaI and CsI, respectively. The dead times assigned to each afterglow inducing event are taken to be the afterglow durations from the C beam in Table \ref{tab:results} for a given detector. Therefore each afterglow event had a 0.28 ms and 2.4 ms duration for the NaI and CsI detectors, respectively. Extrapolating the afterglow particle count rates and associated dead times over an entire orbit, we find that the NaI detector loses approximately 0.02 s to afterglows per orbit while the CsI detector loses approximately 9 seconds. Thus the observing time lost to afterglow in the NaI detector is $\sim$0.01$\times$ that in the CsI detector.

\subsection{BTO Detector Trade Study}
\label{subsect:detectors}

As a requirement for the BTO project, an extensive detector trade study was conducted to determine the optimal scintillator material for the mission. The two predominant materials explored were NaI and CsI. While NaI crystals have a high efficiency and largely linear energy response over moderate energy ranges \citep{Bizarri2011, Valentine1998}, they are also hygroscopic and must be hermetically sealed to maintain performance. The CsI crystals are more space-rugged than NaI and also better fit the energy bandpass observed by BTO. However, CsI is known to exhibit strong afterglow signals \citep{Lecoq2020, Rau2005}. Therefore, the results discussed in Section \ref{subsect:analysis} were crucial for selecting the BTO scintillator material.

Afterglow signals in $\gamma$-ray astronomy missions like BTO appear as false GRB or $\gamma$-ray events and produce false triggers. These triggers complicate software and analysis tools as they must be removed from the true $\gamma$-ray trigger populations. BTO will nominally take data in binned-mode to minimize computational power, data transmission rates to the COSI spacecraft computer, and data downlinking rates. However, when an event triggers in the instrument, BTO will switch to event-by-event mode to improve temporal resolution. Therefore, selecting a scintillator with low afterglow is crucial to prevent false triggers from consuming large fractions of the BTO data budget.

There are several electronics techniques that may be used to eliminate afterglow signals. Since afterglows only occur after a large, saturating deposit of energy is made into the scintillator, the first approach might be to incorporate a set dead-time after each saturated event. The duration of this dead-time would be equal to the afterglow duration for the highest energy particle expected. The drawbacks for this method include creating gaps in the data flow and removing any events which occur rapidly back-to-back---which could include multiple GRB occurring coincidentally in the same time frame, or more probably, a $\gamma$-ray event occurring directly after a particle trapped in the Van Allen radiation belts interacts with a detector. From simulations in Section \ref{subsect:countrates}, we expect an afterglow producing event every $\sim$70 s in the NaI detector and every $\sim$0.7 seconds in the CsI detector. Assigning the C beam afterglow durations from Table \ref{tab:results} as the dead times shows that while the NaI detector loses 0.02 s to afterglow induced dead time, the CsI detector loses a full 9 s over a single orbit. Another approach might be to set the trigger threshold higher than the afterglow amplitude associated with the highest energy particle to ensure BTO does not trigger on afterglow events. A main science requirement for BTO, however, is to provide spectral data down to 30 keV to enable the GRB energy peak to be well modeled. Most bright GRB energy spectra peak between 100--1000 keV \citep{Mallozzi1995}, thus a threshold cut sufficient to remove afterglow signals would not allow BTO to reach a low enough energy to measure the GRB spectra below the turnovers. There are other methods which allow afterglow signals to be removed from the data stream that complicate software. Since an afterglow event is intrinsic to a specific detector and not an external event observable by both detectors, a requirement could be set such that an events is only triggered if it is observed in both BTO detectors. However, the BTO detectors will be placed on opposite sides of the COSI payload, therefore it is likely that some transient events will only be detected by one detector. Therefore, requiring both detectors to observe a triggered event is not a good solution for the BTO mission.

Due to the constraints on bandpass and observable time placed by the strong afterglow signatures in CsI, we determined the NaI scintillator is better suited for the development and performance of BTO.

\section{Conclusions}
\label{sect:conclusion}

We irradiated NaI and CsI detectors with SiPM readouts at the HIMAC beamline in Japan to study afterglow signatures in common $\gamma$-ray scintillators. Two particle beams were used in this study---a carbon beam with 350 MeV/u and a helium beam with 230 MeV/u. The study was motivated by the BTO Student Collaboration Project slotted to fly onboard the NASA funded COSI SMEX mission scheduled to launch in 2027. Conducting a trade-study between NaI and CsI scintillators for usage in a transient $\gamma$-ray detection system in the 30 keV to 2 MeV bandpass was a necessary requirement for the BTO project.

Both the NaI and CsI scintillators exhibited afterglow signatures when exposed to the C and He beamlines. However, the afterglow signal in the CsI detector was much stronger with an afterglow duration 8.6$\times$ and 5.6$\times$ the NaI signal for the C and He beams respectively. Even in the weakest radiation, the afterglow signal was stronger in the CsI detector than for the strongest radiation case with the NaI detector. Since BTO will be observing in the 30 keV to 2 MeV range, afterglows will be frequent and pose a risk for false triggering and unnecessarily high data rates. 

Using the \textbf{MEGAlib} toolkit, we then performed simulations to estimate the background rates with the rectangular, flight-model NaI detector. The expected background rate for the full two-detector BTO system is 114 counts/s. This background rate produces particles with energies large enough to induce afterglows at rates of 0.014 counts/s and 0.7 counts/s in the NaI and CsI detectors, respectively. Assigning the C beam afterglow durations as the detector's afterglow induced dead time, we find the NaI detector only loses 0.02 seconds to afterglows per orbit while the CsI detector loses 9 seconds. Therefore, we determined the NaI scintillator is better suited for a soft $\gamma$-ray mission such as the BTO mission.


\section*{Acknowledgements}

The authors would like to acknowledge Alex Lowell (SSL), Brent Mochizuki (SSL), and Naoki Itoh (TUS) for their support and guidance in developing and executing this experiment.

COSI and BTO are a NASA Small Explorer mission and Student collaboration project supported under NASA contract 80GSFC21C0059. This material is based upon work supported by the National Science Foundation Graduate Research Fellowship under Grant No. DGE 2146752. Hiroki Yoneda acknowledges support by the Bundesministerium für Wirtschaft und Energie via the Deutsches Zentrum für Luft- und Raumfahrt (DLR) under contract number 50 OO 2219.


\bibliography{report} 
\bibliographystyle{jwapjbib} 






\end{document}